\PassOptionsToPackage{table}{xcolor}
\documentclass[acmsmall,screen,nonacm]{acmart}

\usepackage[T1]{fontenc}
\usepackage[utf8]{inputenc}
\usepackage{amsmath}

\usepackage{amssymb}
\usepackage{graphicx}
\usepackage{booktabs}
\usepackage{multirow}
\usepackage{array}
\usepackage{tabularx}
\usepackage{listings}
\usepackage{wrapfig}
\usepackage{tikz}
\usetikzlibrary{arrows.meta,positioning,tikzmark}
\usepackage{pgfplots}
\pgfplotsset{compat=1.18}
\usepackage{placeins}
\usepackage{float}

\usepackage{enumitem}

\usepackage[most]{tcolorbox}

\newtcolorbox{rqanswerbox}{
  colback=gray!10,
  colframe=blue!30!black!20,
  boxrule=0.4pt,
  arc=6pt,
  left=6pt,
  right=6pt,
  top=4pt,
  bottom=4pt,
  enhanced,
  breakable
}

\newcolumntype{Y}{>{\raggedright\arraybackslash}X}
\newcolumntype{C}[1]{>{\centering\arraybackslash}p{#1}}
\newcommand{\sem}{\textsf{S}}
\newcommand{\crash}{\textsf{C}}
\definecolor{codeblue}{HTML}{2563EB}
\definecolor{flowviolet}{HTML}{7C3AED}
\definecolor{alertcoral}{HTML}{F97316}
\definecolor{interestgreen}{HTML}{059669}
\definecolor{softgray}{HTML}{64748B}
\definecolor{palegray}{HTML}{F8FAFC}
\definecolor{bugheader}{RGB}{75,84,99}
\definecolor{bugsubheader}{RGB}{224,228,234}
\definecolor{bugstripe}{RGB}{246,247,249}
\definecolor{bugyellow}{RGB}{255,245,157}

\newcommand{\red}[1]{}

\newcommand{\pc}[1]{}
\definecolor{darkgreen}{RGB}{0,150,80}

\title{QuSema: Detecting Silent Bugs in Quantum Libraries via Quantum-knowledge-enhanced Agents}

\hypersetup{hidelinks}
\usepackage{bookmark}
\graphicspath{{Qusema_arxiv/}{./}{Qusema_arxiv/figure/}{figure/}}

\setcopyright{none}
\author{Yujin Song}
\affiliation{\institution{Nanyang Technological University}\city{Singapore}\country{Singapore}}
\email{SONG0316@e.ntu.edu.sg}
\author{Kaining Zhang}
\authornote{Corresponding authors.}
\affiliation{\institution{Nanyang Technological University}\city{Singapore}\country{Singapore}}
\email{kaining.zhang@ntu.edu.sg}
\author{Qixin Zhang}
\affiliation{\institution{Nanyang Technological University}\city{Singapore}\country{Singapore}}
\email{qixin.zhang@ntu.edu.sg}
\author{Shuai Wang}
\affiliation{\institution{Hong Kong University of Science and Technology}\city{Hong Kong}\country{China}}
\email{shuaiw@cse.ust.hk}
\author{Pingchuan Ma}
\authornotemark[1]
\affiliation{\institution{Zhejiang University of Technology}\city{Hangzhou}\country{China}}
\email{pma@zjut.edu.cn}
\author{Yuxuan Du}
\authornotemark[1]
\affiliation{\institution{Nanyang Technological University}\city{Singapore}\country{Singapore}}
\email{yuxuan.du@ntu.edu.sg}

\begin{document}
\begin{abstract}
Quantum libraries are now critical infrastructure for quantum algorithm development, yet their correctness remains difficult to test. Existing testing techniques mainly rely on failure-based or comparison-based oracles, exposing bugs only when executions fail, violate runtime checks, or disagree with another implementation. Their applicability is limited when suitable execution-based oracles are unavailable, leaving some silent bugs undetected. Such missed bugs can produce incorrect results that propagate into experimental conclusions, simulation studies, and algorithmic designs.
Here we present QuSema, an autonomous testing agent for finding silent bugs in quantum libraries. QuSema uses constraints from quantum semantics and documentation as a source-level semantic oracle to assess whether implementation logic can produce invalid outputs from valid inputs. It operates through an agentic loop that repeatedly inspects library API documentation and source code, reasons about the intended behavior of quantum operations, identifies potential semantic deviations, and validates them by generating executable tests through library APIs. Guided by quantum domain reasoning, QuSema turns high-level behavioral mismatches into concrete, user-triggerable bug reports, enabling it to uncover non-crash defects. We implement QuSema for Qiskit and PennyLane. On a benchmark of 20 historical silent bugs, QuSema achieves higher mean bug relocation counts than Claude Code and DeepSeek Harness, with the DeepSeek configuration outperforming Claude Code at a lower estimated mean benchmark cost. QuSema also discovers 40 previously unknown bugs confirmed by the developers, including 30 silent bugs.
\end{abstract}

\keywords{quantum library testing, quantum computing, silent bugs, coding agents}

\maketitle
\section{Introduction}
\label{sec:introduction}

Quantum computing~\cite{dalzell2025quantumalgorithms,nielsen2010quantum} offers the potential to tackle computational problems beyond the reach of classical methods, with promising applications in quantum chemistry~\cite{mcardle2020quantum}, optimization~\cite{abbas2024quantum}, and machine learning~\cite{cerezo2022challenges}. %
To facilitate the design and refinement of quantum algorithms, quantum software libraries such as Qiskit~\cite{qiskit} and PennyLane~\cite{pennylane} have been developed and widely adopted. %
These libraries provide essential support for building, simulating, and executing quantum algorithms, and have become critical infrastructure in the quantum software ecosystem. However, quantum software is still in its early stage, and these rapidly evolving libraries contain bugs that lead to incorrect or unexpected behaviors~\cite{quantumbugs}. Accordingly, systematic quantum software testing~\cite{dwivedi2024quantumsoftware} is becoming increasingly essential for building a reliable and robust quantum software ecosystem.

Substantial research effort has been devoted to quantum software testing, particularly automated bug detection for quantum libraries~\cite{paltenghi2024survey}. Existing approaches have shown promising results~\cite{kqfuzz,fuzz4all} for \textit{crash bugs}~\cite{rosa2026schrodinger}, where exceptions or program termination provide explicit failure signals.
Beyond crash bugs, quantum libraries also contain \textit{silent bugs}~\cite{zappin2026challenges}, which return incorrect results that violate intended behavior or quantum semantics without explicit failures. For example, a quantum library may return an incorrect commutativity judgment for a pair of quantum operators~\cite{qiskit16164} or apply a quantum circuit optimization method that does not preserve semantics of the original circuit~\cite{qiskit13079}. 
These incorrect but undetected results can propagate through downstream workflows, which in turn mislead algorithm design, lead to erroneous hardware executions, and waste quantum computing resources~\cite{qsimbugs}. To mitigate these risks and improve reliability, quantum software testing should be capable of effectively detecting silent bugs in quantum libraries.

Current approaches to silent bug detection in quantum libraries mainly rely on differential testing~\cite{qdiff,fuzzq,blackwell2024fuzzing,qspe,qite} and metamorphic testing~\cite{morphq,morphqpp,qemi}. Differential testing analyzes the same quantum circuit across different quantum software stacks and checks whether equivalent outputs are observed~\cite{andrews2026survey}. Metamorphic testing instead applies transformations that should preserve circuit semantics, such as inserting identities or replacing a gate with an equivalent decomposition, and checks the expected relations between execution results from the source and transformed programs~\cite{fortunato2024verification}. 
Both approaches use black-box test oracles~\cite{rigger2022intramorphic}  to validate expected relations between execution results~\cite{gao2026dbmsfuzzing}.
While these approaches have successfully detected silent bugs in quantum libraries~\cite{qdiff,fuzzq,blackwell2024fuzzing,qspe,qite,morphq,morphqpp,qemi}, their reliance on black-box test oracles introduces two fundamental limitations.

\textbf{(L1).} \textit{Limited Availability of Expected Relations Between Execution Results}. Some quantum library functionalities lack either semantically corresponding implementations across libraries for differential testing or circuit transformations with predictable output relations for metamorphic testing. For example, PennyLane supports the resource estimation functionality~\cite{pennylanedocs} for predefined algorithms, while Qiskit has no counterpart~\cite{qiskitdocs}, and semantics-preserving transformations may still change resource quantities such as gate count or circuit depth. Consequently, approaches that employ black-box test oracles cannot provide complete coverage of quantum library functionality.

\textbf{(L2).} \textit{Ineffective Triggering Without Semantic Guidance}. Many quantum software functionalities involve complex mathematical semantics~\cite{salam2024quantumchallenges}, and their bugs are triggered only under specific semantic conditions~\cite{roadmap}. Such conditions may involve specific parameter values, quantum operations, or circuit layouts~\cite{upbeat}. Generic program generation may require extensive exploration to trigger these conditions~\cite{li2026methodological}, limiting the effectiveness of black-box testing.

To address these limitations, we develop QuSema, an automated agent for detecting silent bugs in quantum libraries by auditing \textbf{Qu}antum \textbf{Sema}ntics.
Its design follows our data flow analysis, which localizes a silent bug to the library API that first transforms a valid input into an invalid output. Our study of reported bugs in Qiskit and PennyLane provides supporting evidence for this localization. QuSema therefore targets such API-local semantic deviations by directly analyzing the relevant implementation logic.
More precisely, QuSema uses a \emph{source-level semantic {criterion}} that identifies potential defects when implementation logic deviates from quantum semantics or documentation. QuSema then employs large language models (LLMs)~\cite{wei2025equibench,mikuriya2025qcoder} to apply this {criterion} and identify candidate defects. It validates these candidates through execution and checks whether they can be triggered through documented library APIs. This design brings quantum semantics into source-level correctness reasoning and extends silent bug detection beyond cases covered by existing execution-based relations.

We systematically evaluate QuSema on Qiskit and PennyLane, two widely used quantum
libraries. On a benchmark of 20 historical silent bugs, QuSema achieves substantially higher mean bug relocation counts than Claude Code~\cite{anthropicclaudecode} and DeepSeek Harness~\cite{deepseek2026dsh} using the corresponding LLM backends. With DeepSeek, QuSema also outperforms Claude Code at a lower estimated mean benchmark cost. Beyond historical benchmarks, applying QuSema to real-world quantum libraries uncovers 40 previously unknown bugs confirmed by developers, including 30 silent bugs.  The identified bugs span quantum program execution, semantic analysis, and library infrastructure. Notably, five of these silent bugs affect functionalities that lack either comparable implementations or suitable circuit transformations with known output relations. Such bugs fall beyond the detection scope of existing differential and metamorphic testing methods.
These results show the value of grounding code inspection in quantum semantics for silent bug detection in quantum libraries. 

In summary, our main contributions can be summarized in threefold: 

\noindent$\bullet$ We characterize silent bugs in quantum libraries and show that all examined cases are API-local, leaving semantic evidence in implementation logic and documentation. This defines a well-scoped and prevalent bug class for source-level detection with quantum semantics.

\noindent$\bullet$ {We propose a source-level semantic {criterion} that assesses implementation logic against constraints from quantum semantics and documentation. Based on this {criterion}, we develop QuSema, an automated testing agent that applies the {criterion} using LLMs, providing a new paradigm for silent bug detection in quantum libraries.}

\noindent$\bullet$ We demonstrate the effectiveness of QuSema on 20 historical silent bugs through comparisons with coding agents and ablation studies. In addition, QuSema discovers 40 previously unknown bugs in Qiskit and PennyLane that are confirmed by developers, including 30 silent bugs, 5 of which affect functionality beyond the coverage of existing differential and metamorphic testing.

\section{Background}
\label{sec:background}

This section provides the background on quantum computing, quantum library structure, and existing bug detection approaches. In Sec.~\ref{sec:quantum-computing-basics}, we review basic concepts in quantum computing. In Sec.~\ref{sec:api-callable}, we describe the functionality and software structure of quantum libraries. Finally, in Sec.~\ref{sec:background-bug-tools}, we review existing approaches to silent bug detection in quantum libraries.

\subsection{Basic Concepts in Quantum Computing}\label{sec:quantum-computing-basics}

\textbf{Qubits and Quantum States.}
Quantum information is represented by quantum states, with the qubit as the fundamental unit~\cite{nielsen2010quantum}. In the computational basis $\{|0\rangle,|1\rangle\}$, where $|0\rangle=(1,0)^T$ and $|1\rangle=(0,1)^T$, a pure qubit state can be written as $|\psi\rangle=\alpha|0\rangle+\beta|1\rangle$, where $\alpha,\beta\in\mathbb{C}$ satisfy $|\alpha|^2+|\beta|^2=1$. For $n$ qubits, the basis states are formed through tensor products as $|x\rangle=|x_1\rangle\otimes\cdots\otimes|x_n\rangle$ for each $x_i\in\{0,1\}$, which span a $2^n$-dimensional Hilbert space. An $n$-qubit pure state can be expressed as $|\psi\rangle=\sum_{x\in\{0,1\}^n}\alpha_x|x\rangle$, where $\sum_x|\alpha_x|^2=1$.

\textbf{Quantum Gates.}
Quantum gates act on quantum states through unitary operators. Up to a global phase, a single-qubit unitary can be written as $U(\theta,\varphi,\lambda)=\left[\begin{smallmatrix}\cos(\theta/2)&-e^{i\lambda}\sin(\theta/2)\\ e^{i\varphi}\sin(\theta/2)&e^{i(\varphi+\lambda)}\cos(\theta/2)\end{smallmatrix}\right]$. Common single-qubit gates include $H=\frac{1}{\sqrt{2}}\left[\begin{smallmatrix}1&1\\1&-1\end{smallmatrix}\right]$, the Pauli gates $X=\left[\begin{smallmatrix}0&1\\1&0\end{smallmatrix}\right]$, $Y=\left[\begin{smallmatrix}0&-i\\i&0\end{smallmatrix}\right]$, $Z=\left[\begin{smallmatrix}1&0\\0&-1\end{smallmatrix}\right]$, $S=\left[\begin{smallmatrix}1&0\\0&i\end{smallmatrix}\right]$, and Pauli rotations $R_P(\theta)=e^{-i\theta P/2}=\cos(\theta/2)I-i\sin(\theta/2)P$ for $P\in\{X,Y,Z\}$.
Controlled gates apply an operation conditionally. For a single-qubit target,
${\rm CU}(\theta,\varphi,\lambda,\gamma)=|0\rangle\langle0|\otimes I+e^{i\gamma}|1\rangle\langle1|\otimes U(\theta,\varphi,\lambda)$,
which gives CNOT for $U=X$ and $\gamma=0$.

\textbf{Quantum Circuits.}
A quantum circuit represents quantum operations on qubit wires. For example, a single-qubit unitary $U$ is represented as
\begin{tikzpicture}[baseline=-0.6ex]
    \draw (0,0) -- (1.0,0);
    \node[draw,fill=white,minimum width=0.4cm,minimum height=0.3cm,
          inner sep=1pt] at (0.5,0) {$U$};
\end{tikzpicture},
and its controlled version as
\begin{tikzpicture}[baseline=-0.6ex]
    \draw (0,0.15) -- (1.0,0.15);
    \draw (0,-0.15) -- (1.0,-0.15);
    \fill (0.5,0.15) circle (1.5pt);
    \draw (0.5,0.15) -- (0.5,-0.04);
    \node[draw,fill=white,minimum width=0.4cm,minimum height=0.3cm,
          inner sep=1pt] at (0.5,-0.15) {$U$};
\end{tikzpicture},
where the filled dot denotes the control qubit.
For input $|\psi\rangle$ and gates $U_1,\ldots,U_k$ applied left to right, the output is $|\phi\rangle=U_k\cdots U_1|\psi\rangle$. Their arrangement determines circuit depth and gate count. Different circuit structures may implement the same computation, enabling semantics-preserving circuit transformation and compilation.

\subsection{Functionality and Software Structure of Quantum Libraries}
\label{sec:api-callable}

\textbf{Quantum Library Functionality.}
Quantum libraries provide functionality in three main categories: \emph{Quantum programming language}, \emph{Compilation}, and \emph{Execution environment}~\cite{quantumbugs}. The first provides quantum and classical abstractions, such as gates, circuits, matrices, and tensors, together with domain-specific abstractions for quantum algorithms. Compilation supports intermediate representations, circuit optimization, and code generation. The execution environment provides hardware interfaces, simulators, and quantum state evaluation, including measurement and error mitigation. Auxiliary components support testing, visualization, and software integration. Individual libraries implement different parts of this functionality.

\textbf{Library APIs.}
Quantum software libraries expose their functionality through APIs in a host programming language~\cite{upadhyay2025quantumrepositories}. In Qiskit and PennyLane, documented user-facing APIs are primarily Python functions and methods. Their official documentation specifies intended usage, accepted inputs, expected outputs, and relevant behavioral constraints. These APIs provide the interface through which users construct quantum programs and access library functionality.

\textbf{Code Segments.}
A library API is typically implemented by one or more functions or methods in the source code. Following prior work~\cite{coverup}, we refer to each function or method as a \textit{code segment}. A library API may invoke multiple code segments, while the same code segment may participate in multiple APIs. Code segments may also contain local documentation, such as docstrings or source comments. For segments that do not correspond to documented API entries, this local documentation is not part of the official API documentation.

\subsection{Existing Approaches to Bug Detection in Quantum Libraries}
\label{sec:background-bug-tools}

Bugs can manifest as explicit crashes such as exceptions and resource exhaustion, or as silent bugs that produce incorrect behavior without interrupting execution~\cite{rosa2026schrodinger}. An empirical study~\cite{quantumbugs} shows that quantum-specific bugs frequently manifest as incorrect outputs rather than crashes. Accordingly, silent bug detection in quantum libraries has received substantial attention. Existing methods in this context primarily fall into two categories.

\textbf{Differential Testing.}
Quantum differential testing detects inconsistencies by comparing corresponding executions across different quantum software stacks~\cite{andrews2026survey}. Concretely, QDiff~\cite{qdiff} and QITE~\cite{qite} derive comparable programs through program transformations. Blackwell et al.~\cite{blackwell2024fuzzing} and FuzzQ~\cite{fuzzq} employ fuzzing to generate test programs, guided respectively by predefined grammars and constraints from a formal model. QSPE~\cite{qspe} systematically enumerates program variants from skeletal programs and compares their final state vectors.

\textbf{Metamorphic Testing.}
Quantum metamorphic testing applies transformations that should preserve circuit semantics and checks whether the resulting executions satisfy expected relations~\cite{fortunato2024verification}. Specifically, MorphQ~\cite{morphq} derives transformed programs using quantum metamorphic relations, while MorphQ++~\cite{morphqpp} extends these transformations to additional circuit representations and transpilation settings. QEMI~\cite{qemi} constructs equivalent program variants by eliminating code that does not affect selected inputs in quantum programs with control flow. These methods exercise circuit construction, compilation, and execution, involving functionality from the three categories introduced in Sec.~\ref{sec:api-callable}. Moreover, their applicability to individual functionalities depends on whether corresponding executions or suitable circuit transformations can be constructed.

Differential and metamorphic testing rely on black-box test oracles that evaluate relations between execution results. Their applicability therefore depends on the availability of comparable executions or suitable metamorphic relations, leaving some quantum library functionalities difficult to test. In contrast, QuSema derives semantic oracles directly from source-level implementation logic, enabling silent bug detection without requiring such predefined relations.

\textbf{LLM-based Testing}. Recent studies have explored large language models (LLMs) for test generation, input construction, fuzzing, and debugging~\cite{titanfuzz,fuzzgpt,fuzz4all,dfuzz,fuzzllm,whitefox,llmfuzzdriver,coverup,qbuglm}. These methods primarily use LLMs to generate test cases and expose failures through program execution, without directly reasoning about semantic defects in library implementations. In contrast, QuSema uses the LLM to analyze implementation logic and identify potential semantic deviations. These candidates are subsequently verified and checked for triggerability through documented library APIs.

\section{Current Limitations and Semantic Evidence for Quantum Silent Bug Detection}
\label{sec:characterization}

In this section, we analyze the limitations of existing silent bug detection methods in quantum libraries and explore source-level semantic evidence as a potential basis for broader and more efficient bug detection. For clarity, we first analyze two limitations of existing approaches employing black-box test oracles in Sec.~\ref{sec:motivation-incomplete-functionality}, namely \textbf{(L1)} incomplete functionality coverage due to unavailable expected relations between execution results and \textbf{(L2)} inefficient bug triggering without semantic guidance. {We then analyze in Sec.~\ref{sec:api-source-evidence} how data flow can be used to localize semantic deviations to individual library APIs and examine reported bugs to support this analysis.} Finally, in Sec.~\ref{sec:challenges}, we discuss the implications and challenges of using this evidence for reliable silent bug detection.

\subsection{Two Limitations of Existing Approaches}
\label{sec:motivation-incomplete-functionality}

\begin{figure}[t]
\centering
\includegraphics[width=\linewidth]{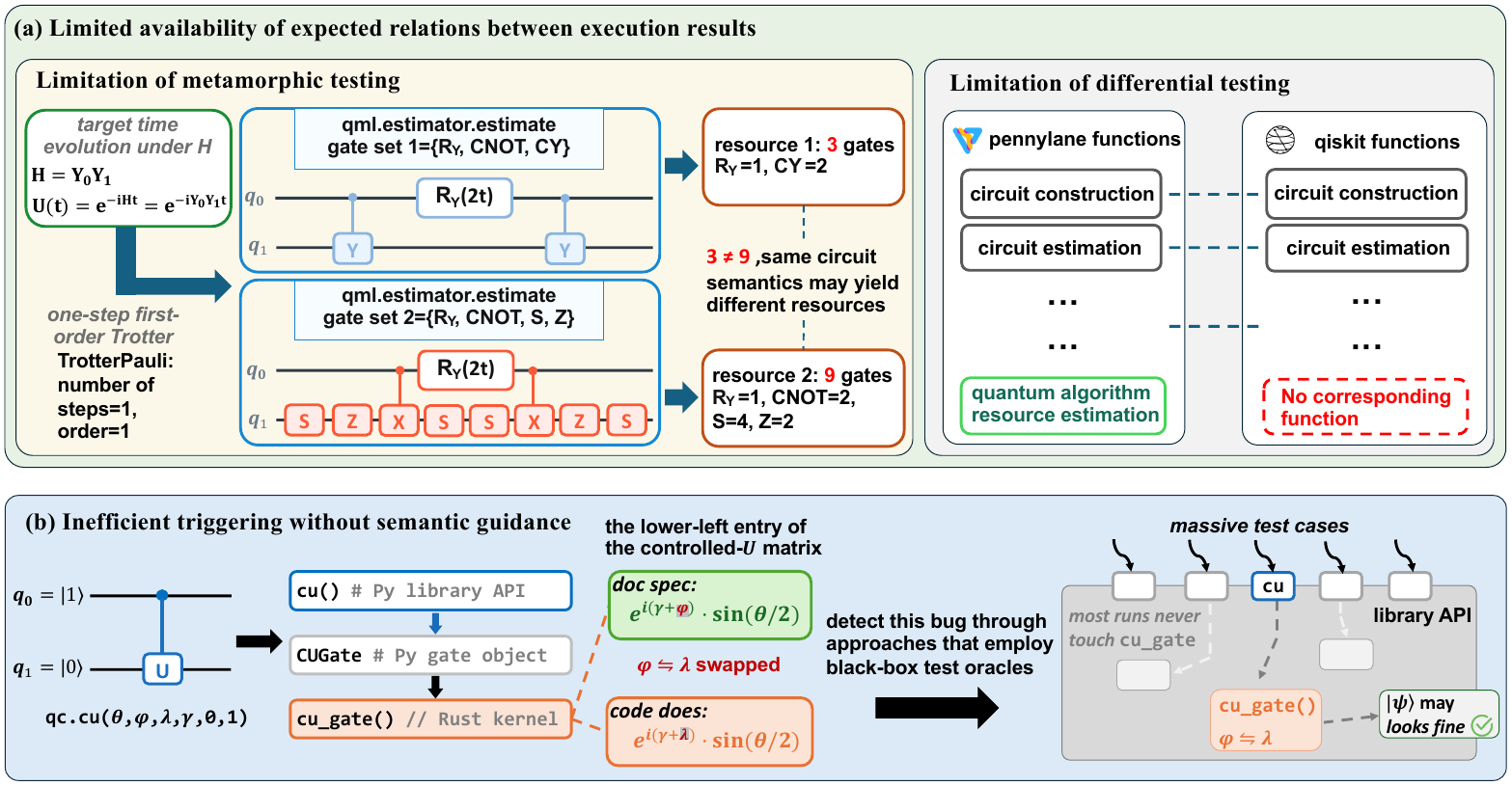}
\Description{Two-panel figure on the limitations of black-box test oracles. Panel (a) shows two equivalent circuits for the same computation that receive different resource counts, and a resource estimation functionality without a cross-library counterpart. Panel (b) shows a bug in the implementation of the CU gate that is exposed only under specific semantic conditions.}
\caption{Limitations of approaches employing black-box test oracles.
(a) Limited availability of expected relations between execution results, where equivalent circuits can have different resource counts and some higher-level resource estimation functionality lacks a cross-library counterpart.
(b) Inefficient triggering, where a bug when implementing the CU gate is exposed only under specific semantic conditions. }
\label{fig:execution-first-limitations}
\end{figure}

Existing approaches for detecting silent bugs in quantum libraries, particularly differential and metamorphic testing introduced in Sec.~\ref{sec:background-bug-tools}, mainly rely on black-box test oracles~\cite{rigger2022intramorphic}. They have shown effectiveness for functionalities such as circuit execution and compilation, yet two limitations defined in Sec.~\ref{sec:introduction} (i.e., \textbf{(L1)} and \textbf{(L2)}) remain. For some quantum library functionalities, the expected relations between execution results cannot be established. This occurs when cross-library counterparts are unavailable or circuit-level transformations do not yield predictable output relations. Moreover, many silent bugs are triggered only under specific quantum semantic conditions, making generic program generation inefficient. Below, we elaborate on these two limitations with representative examples in Fig.~\ref{fig:execution-first-limitations}.

\textbf{(L1) Limited Availability of Expected Relations Between Execution Results.}
For some functionalities in quantum libraries, the expected relations between execution results needed for differential or metamorphic testing cannot be established. Here, we use resource estimation as a representative example to illustrate the limitations of both metamorphic and differential testing. As shown in Fig.~\ref{fig:execution-first-limitations}(a), PennyLane provides the \texttt{qml.estimator.estimate} API to estimate the gate count required to implement a quantum circuit or operator under a specified gate set.

For metamorphic testing, consider Trotterized time evolution under the Hamiltonian $H=Y_0Y_1$. The same functionality reports 3 and 9 gates for two equivalent circuit representations using the gate sets $\{R_Y,\mathrm{CNOT},\mathrm{CY}\}$ and $\{R_Y,\mathrm{CNOT},S,Z\}$, respectively. Because the gate count is not preserved across semantics-preserving circuit transformations, the expected relation between the two outputs is unavailable as a metamorphic oracle.

For differential testing, the same resource estimation functionality presents a different limitation. PennyLane supports resource estimation for predefined quantum algorithms such as the illustrated Trotter decomposition, while Qiskit does not provide a corresponding functionality. Consequently, no cross-library output is available to construct a differential oracle.

\textbf{(L2) Inefficient Triggering Without Semantic Guidance.}
Approaches based on black-box test oracles rely on testing generated programs to expose faulty behavior. In practice, many silent bugs in quantum libraries are triggered under specific semantic conditions involving parameter values, quantum states, or circuit structures. Generic testing may require substantial exploration to reach the defective implementation under these conditions. 

We use Qiskit \#13118~\cite{qiskit13118} in Fig.~\ref{fig:execution-first-limitations}(b) as a representative case. As shown on the left side of the figure, the Python library API \texttt{qc.cu} is designed to implement the gate CU. Its call path passes through \texttt{CUGate} and reaches the Rust kernel \texttt{cu\_gate}, which incorrectly swaps parameters $\phi$ and $\lambda$. A test exposes this bug only if it reaches \texttt{cu\_gate} with $\phi \neq \lambda$ and $\sin(\theta/2) \neq 0$. Generic testing may require repeated executions to meet these conditions, leading to inefficient bug triggering as shown on the right side of Fig.~\ref{fig:execution-first-limitations}(b).

These examples reflect two key limitations of existing approaches that employ black-box test oracles for quantum libraries. These limitations motivate alternative approaches. 

\subsection{Motivating Study: Source-Level Semantic Evidence}
\label{sec:api-source-evidence}

In this section, we analyze how silent bugs arise along the data flow of quantum programs and how their origins can be identified from source-level semantic evidence. Specifically, we first characterize silent bugs as semantic deviations that produce invalid outputs from valid inputs and localize each deviation to the first library API that introduces it. We then use a representative bug to illustrate how inconsistencies between implementation logic and quantum semantics or documentation provide evidence for identifying such defects.

\textbf{Characterization of Silent Bugs.}
{We characterize a silent bug as an implementation defect that produces an invalid output from a valid input without an explicit failure signal. To assess validity, we consider both library documentation and quantum semantics. Documentation describes the intended behavior of APIs but may omit relevant constraints. Quantum semantics specify the mathematical meaning of quantum objects and operations, including admissible inputs and expected behavior, as discussed in Sec.~\ref{sec:motivation-incomplete-functionality}. Therefore, an input or output is valid if it satisfies the relevant quantum semantics and does not violate the documentation, and invalid otherwise. At the code segment level, we additionally consider applicable software semantics, since local implementation logic may manipulate objects not fully constrained by quantum semantics, and violations of these software semantics may ultimately manifest as quantum-semantic deviations.

\textbf{Localization of Semantic Deviations.}
A quantum program may invoke multiple library APIs, with the output of one call serving as input to subsequent calls. Within this flow, we treat each library API as a data transformation unit from its received input to its resulting output. For an execution exhibiting a silent bug, the first library API that transforms a valid input into an invalid output introduces the semantic deviation and is identified as its origin. Earlier calls preserve validity, while later calls may only propagate or expose the resulting incorrect behavior. This data flow view localizes the semantic deviation to an individual library API.
To examine whether reported bugs provide counterexamples to this localization, we screened 3,184 bug reports from Qiskit and PennyLane using Codex with GPT-5.6-Sol (Ultra). We specifically searched for cases in which each API behaves correctly in isolation, but a permitted composition of APIs produces incorrect behavior. The screening surfaced two candidate counterexamples, Qiskit~\#4369~\cite{qiskit4369} and PennyLane~\#534~\cite{pennylane534}. In both cases, an output accepted as valid by one component is rejected by a subsequent API, producing an explicit exception instead of a silent incorrect result. Thus, no silent bug counterexample to the API localization criterion was found among the screened reports. Accordingly, we scope this work to \emph{API-local silent bugs}, whose semantic deviation can be localized to the first library API that transforms a valid input into an invalid output.

\begin{figure}[t]
\centering
\includegraphics[width=\linewidth]{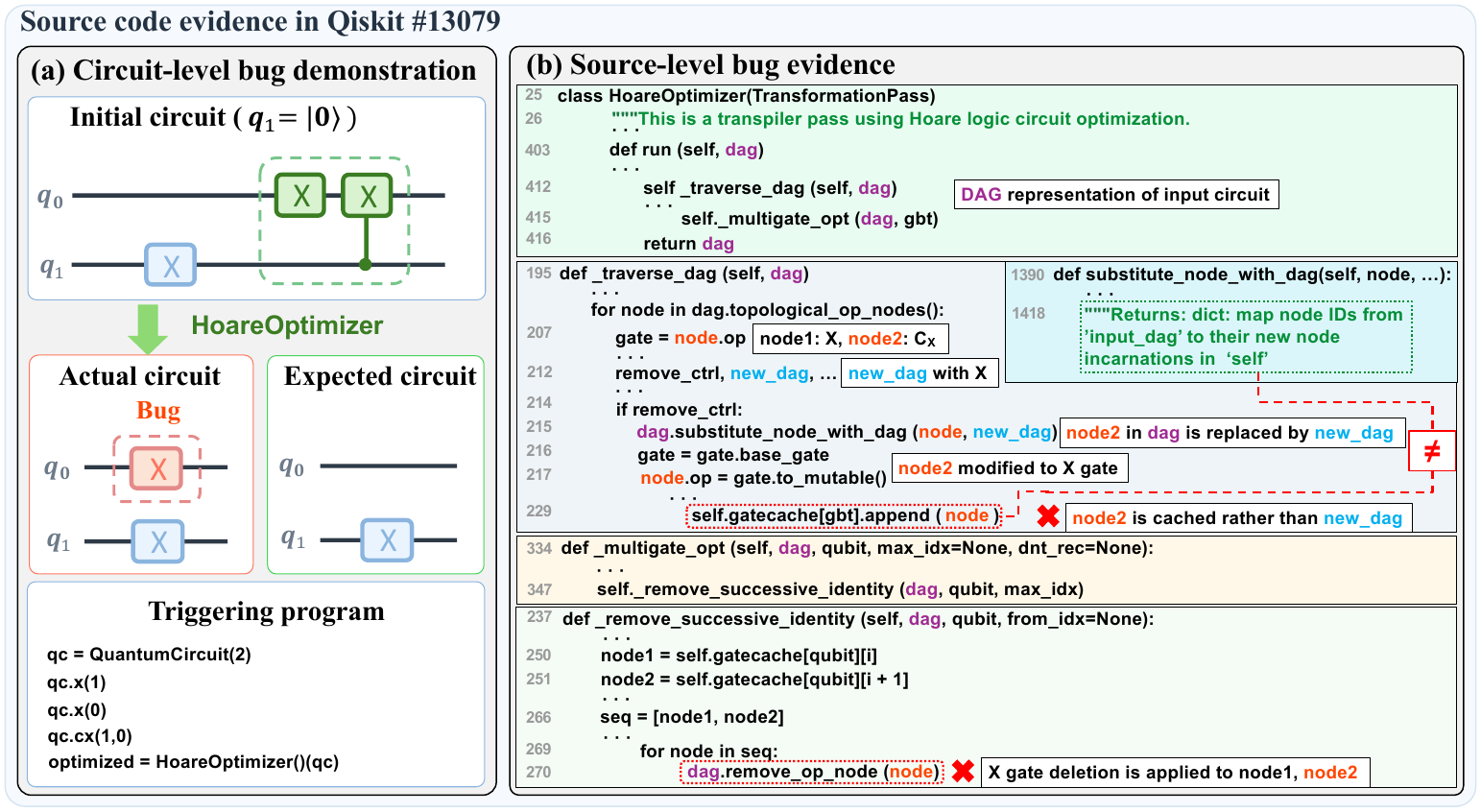}
\Description{Two-panel figure for Qiskit issue 13079 in HoareOptimizer. Panel (a) shows the triggering program together with the actual and expected output circuits; the actual circuit contains an extra X gate on qubit q0. Panel (b) shows the source code, where the original node is cached after being replaced, so a later cancellation leaves the replacement X gate in the circuit.}
\caption{Circuit-level manifestation and source-level semantic evidence of the \texttt{HoareOptimizer} bug in Qiskit \#13079~\cite{qiskit13079}. (a) The triggering program and actual and expected circuits, showing an extra $X$ on $q_0$. (b) The implementation caches the original node after replacing it, so subsequent cancellation leaves the replacement $X$ in the circuit. This violates quantum circuit semantics and is inconsistent with the documented node replacement.}
\label{fig:source-evidence-13079}
\end{figure}

\textbf{Inconsistencies with Quantum Semantics and Documentation.}
Both forms of semantic evidence are illustrated by Qiskit \#13079~\cite{qiskit13079} as shown in Fig.~\ref{fig:source-evidence-13079}. The bug affects the Qiskit API \texttt{HoareOptimizer}, described in the documentation of \texttt{hoare\_opt.py}~\cite{qiskit120hoareopt} for circuit optimization. As shown in Fig.~\ref{fig:source-evidence-13079}(a), when the control qubit is $\lvert 1\rangle$, the CX gate can be simplified to an $X$ gate on the target. This $X$ gate should cancel with the adjacent $X$ gate because $X^2=I$, whereas the implementation leaves an extra $X$ gate in the circuit. The incorrect simplification arises from the internal function \texttt{\_traverse\_dag}, as shown in Fig.~\ref{fig:source-evidence-13079}(b). That is, after updating the circuit using \texttt{substitute\_node\_with\_dag}, the subsequent code should process the replacement node but instead continues to process the original node. This behavior violates the quantum circuit semantics that the optimization should preserve and provides direct evidence of the bug. This defect is also exposed by the documentation of \texttt{substitute\_node\_with\_dag} in \texttt{dagcircuit.py}~\cite{qiskit120dagcircuit}, as the function replaces a node and returns a mapping to the newly inserted nodes. Continuing to process the original node after replacement is inconsistent with the documented behavior. Both quantum semantics and documentation provide semantic evidence of the same defect.

{The above data flow analysis motivates API-local defect analysis, and the issue study provides supporting evidence from reported bugs.} The Qiskit case in Fig.~\ref{fig:source-evidence-13079} further illustrates how such deviations can be identified from source-level semantic evidence without relying on expected relations between execution results, addressing \textbf{(L1)}. Besides, the associated semantic conditions can guide the construction of triggering inputs, helping mitigate \textbf{(L2)}.

\subsection{Implications and Challenges}
\label{sec:challenges}

The motivating study in Sec.~\ref{sec:api-source-evidence} suggests that source-level semantic evidence can help address \textbf{(L1)} and \textbf{(L2)} in existing approaches to silent bug detection in quantum libraries. Turning this insight into a practical testing method requires systematically identifying and validating such evidence, which introduces two key challenges.

\textbf{(C1) Incorporating Quantum Semantics into Source-Level Correctness Reasoning.}
As discussed in Sec.~\ref{sec:api-source-evidence}, discovering silent bugs requires analyzing source code, interpreting documentation, and reasoning with quantum domain knowledge. Documentation can encode semantic constraints on API behavior, whose interpretation may itself require such knowledge, as illustrated by the inconsistency between documentation and implementation logic in Fig.~\ref{fig:source-evidence-13079}. Since these constraints often involve mathematical and domain-specific concepts, the first challenge is to integrate relevant quantum semantics into correctness reasoning over library implementation logic.

\textbf{(C2) Distinguishing Internal Defects from Bugs Exposed Through Valid Library API Use.}
A source-level semantic defect may be unreachable through documented library
APIs because higher-level logic prevents the defective code from being
executed. We exclude such unreachable defects and consider
only those that can arise during valid API use. The second challenge is
therefore to determine whether an identified defect is triggerable through a
documented library API.

\section{QuSema: Semantics-Driven Detection of Silent Bugs in Quantum Libraries}
\label{sec:method}

This section presents QuSema, a framework for detecting silent bugs in quantum libraries using semantic evidence from implementations. Inspired by prior work on large language models (LLMs) reasoning about program semantics~\cite{wei2025equibench} and quantum programming~\cite{mikuriya2025qcoder}, 
QuSema leverages the reasoning and coding abilities of LLMs to identify semantic deviations by analyzing implementation logic against quantum semantics and documentation.

There are two critical issues in completing such a reasoning process. First, it is unknown how to assign an appropriate analysis scope. More concretely, excessive context may contain irrelevant information that interferes with LLM reasoning~\cite{blinn2024typedholes}, whereas insufficient context may omit information required for correct analysis~\cite{sheng2026fuzzingbrainv2}. Each code segment forms a complete data flow unit within a library API, allowing a silent bug to be localized to the first segment that introduces a semantic deviation. Subsequent segments may propagate or expose it, as illustrated in Fig.~\ref{fig:source-evidence-13079}. Second, analyzing the target code segment may require implementation logic defined outside its boundaries. To address both issues in quantum software testing, QuSema adopts a structured analysis around each code segment together with relevant documentation and call relations, which collectively constitute a \emph{contextual unit}. QuSema analyzes each contextual unit for semantic defects and then determines whether the identified defects can be triggered through library APIs. In this way, QuSema keeps a focused analysis scope while incorporating additional context when needed, which supports more effective LLM reasoning.

\subsection{Overview of QuSema}

QuSema follows a three-stage workflow for discovering and validating silent bugs, as illustrated in Fig.~\ref{fig:overview}. 
{Its key principle is to use a source-level semantic {criterion} to guide defect analysis, and employ LLMs to identify semantic deviations in implementation logic and assess their triggerability through valid library API use.
}

The workflow begins with Stage I, \emph{Contextual Unit Preparation}, which establishes a focused scope for semantic analysis. QuSema uses code segments as the primary analysis units and associates them with documentation that constrains their intended behavior and call relations that provide access to related implementation logic when needed. The resulting contextual units reduce unrelated source context while preserving information required to interpret the behavior of each segment.

Next, in Stage II, \emph{Semantic Defect Detection}, 
{QuSema applies the source-level semantic {criterion} to identify potential semantic violations in each contextual unit.} Each suspected violation is converted into a testable defect hypothesis with predicted erroneous behavior and validated through execution. Hypotheses supported by the observed results are retained as candidate defects. 
{This stage addresses \textbf{(C1)} by incorporating quantum domain knowledge into source-level correctness reasoning.}

Finally, in Stage III, \emph{Library API Triggering and Bug Validation}, QuSema determines whether a verified source-level defect can be reproduced through valid use of a documented library API. It searches for API entries that can reach the defective implementation, constructs user-level inputs that may preserve the triggering condition identified in Stage II, and verifies whether the same semantic deviation is exposed at the API level. The reverse chain in Fig.~\ref{fig:overview}(c) provides additional guidance for this search. Reproduced cases are then classified according to input validity and API documentation. This stage addresses \textbf{(C2)} by distinguishing internal implementation defects from user-triggerable library bugs.

Across these stages, QuSema progresses from library source code to focused contextual units, verified semantic defects, and finally user-triggerable bugs. We next detail these stages in Secs.~\ref{sec:design-slicing}, \ref{sec:design-exploration}, and \ref{sec:design-triggering}, respectively.

\begin{figure}[t]
\centering
\includegraphics[width=\linewidth]{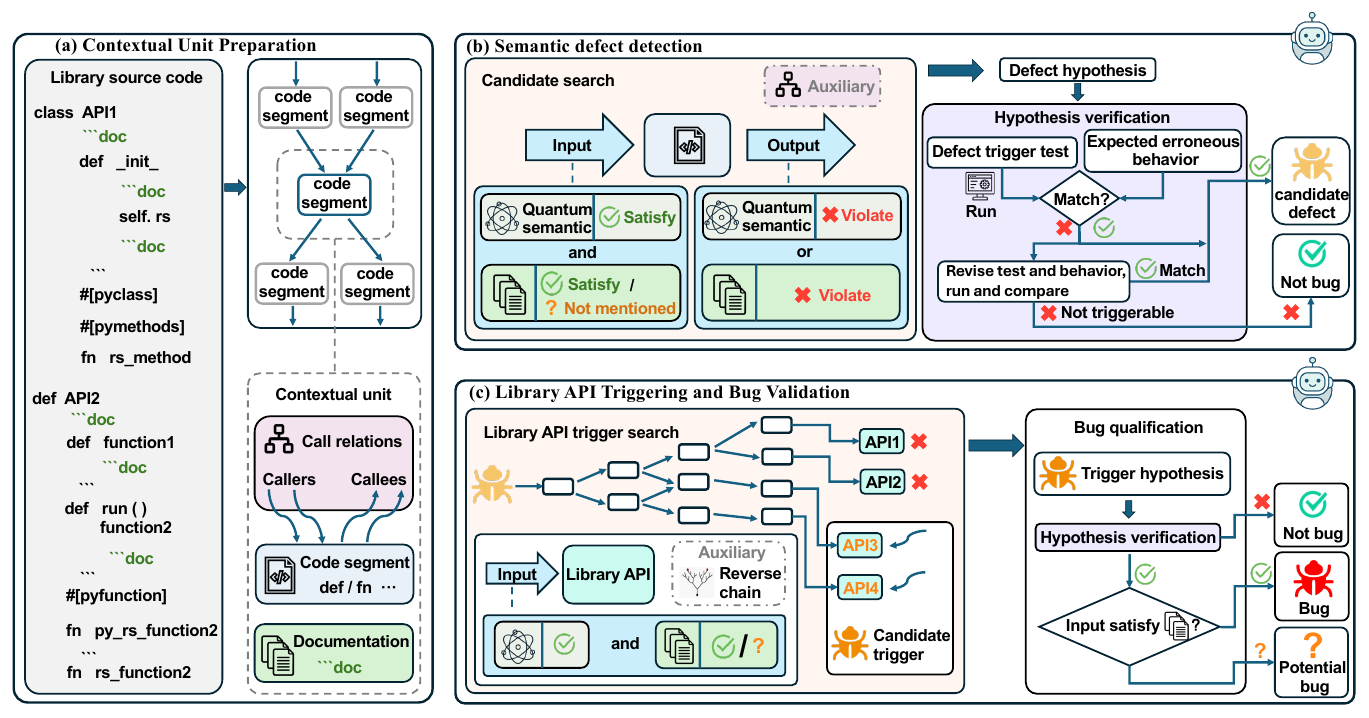}
\Description{Three-stage workflow diagram of QuSema. Stage (a) extracts code segments from library source code and combines them with call relations and documentation into contextual units. Stage (b) searches each unit for semantic violations, formulates defect hypotheses, and runs tests to obtain candidate defects. Stage (c) searches for documented APIs that expose each candidate, using the reverse call chain as guidance, verifies the triggers, and classifies results as bugs, potential bugs, or not bugs.}
\caption{Workflow of QuSema for detecting silent bugs in quantum libraries.
(a) Contextual Unit Preparation extracts code segments from library source code and combines them with call relations and documentation to form contextual units. (b) Semantic Defect Detection searches each contextual unit for semantic violations, formulates defect hypotheses and executes tests to produce candidate defects. (c) API Triggering and Bug Validation searches for documented APIs that expose the candidate defects, uses the reverse chain as auxiliary guidance, verifies the triggers, and classifies the results as \textit{bugs}, \textit{potential bugs}, or \textit{not bugs}.}
\label{fig:overview}
\end{figure}

\subsection{Contextual Unit Preparation}
\label{sec:design-slicing}

Stage I establishes the scope of semantic analysis. An entire source file may contain substantial logic unrelated to the behavior under examination, while an isolated function may omit dependencies needed to interpret that behavior. Following the localization in Sec.~\ref{sec:api-source-evidence}, QuSema analyzes individual code segments and retrieves related information as needed. Specifically, a contextual unit contains the target code segment, documentation that constrains its intended behavior, and call relations that provide access to relevant implementation logic. This design keeps the initial reasoning scope focused while preserving access to dependencies that may affect the semantic behavior of the segment. For example, the Qiskit~2.4.1 Rust file \texttt{standard\_gates\_commutations.rs}~\cite{qiskit241commutations} contains 7,778 lines covering gate metadata, commutation tables, and helper logic. Analyzing the complete file introduces substantial unrelated context, whereas analyzing an extracted function alone may omit referenced tables or surrounding logic required to determine whether its commutation behavior is correct. QuSema therefore constructs contextual units through code segment extraction, call relation analysis, and documentation association.

\textbf{Code Segment Extraction.}
QuSema extracts individual code segments from the quantum library source code shown on the left side of Fig.~\ref{fig:overview}(a). For Python, QuSema uses source positions from the standard \texttt{ast} module~\cite{pythonast} to extract module-level functions and class methods, while nested functions remain within their enclosing function. For Rust, regular expressions identify module-level functions and functions or methods in \texttt{impl} blocks, with brace depth tracking determining their boundaries. Functions exposed through PyO3 are also included~\cite{pyo3}. Each segment retains its attached Python decorators or Rust attributes and records its qualified name, signature, source file, and line range. QuSema further partitions the few Rust functions exceeding 1,000 lines while retaining the enclosing function context.

\textbf{Call Relation Analysis.}
A code segment may depend on implementation logic defined elsewhere. QuSema records these dependencies for inspection without expanding the initial scope to the entire source file. The Call relations component in Fig.~\ref{fig:overview}(a) records callers and callees~\cite{code2graph}, together with class references, variable definitions, and cross-language relations. Source locations are retained for inspecting related code. For Python, call targets are resolved from call expressions and import bindings. For Rust, QuSema uses call sites, enclosing \texttt{impl} types, and qualified names. Referenced constants, class attributes, and tables from the same file are included when available. PyO3 export attributes and module registrations connect Rust implementations to their corresponding Python interfaces~\cite{pyo3}. Unresolved calls are retained for later inspection. For partitions of large Rust functions, QuSema also records definitions and updates of local variables outside the partition.

\textbf{Documentation Association.}
Semantic analysis also requires evidence of the implementation's intended behavior. QuSema therefore associates each code segment with relevant documentation, as illustrated in Fig.~\ref{fig:overview}(a). Python docstrings are obtained from syntax tree nodes, while Rust documentation is collected from documentation comments (\texttt{///} and \texttt{//!}) and the \texttt{\#[doc = ...]} attribute~\cite{rustdoccomments}. For Python class methods, class-level documentation is also included when it specifies relevant behavior. Documentation from the constructor, resolvable base classes, and directly related classes is included when available. For Rust segments, QuSema resolves documentation from the corresponding Python interface through exported names and containing types, using related \texttt{\#[pyclass]} types as fallback. Partitions of large Rust functions retain the enclosing function documentation and relevant gate class documentation. Documentation sources and missing documentation are explicitly recorded.

The design in Stage I of QuSema organizes library code into compact contextual units with explicit links to supporting information. This structure limits irrelevant context while preserving access to dependencies that may affect code behavior. The subsequent semantic defect detection stage can then begin from a focused scope and retrieve additional context when needed.

\subsection{Semantic Defect Detection}

\label{sec:design-exploration}

After contextual unit preparation, Stage II performs \emph{Semantic Defect Detection}, as shown in Fig.~\ref{fig:overview}(b). Based on the characterization of silent bugs in Sec.~\ref{sec:api-source-evidence}, QuSema applies a \emph{source-level semantic criterion} that identifies a potential silent bug when implementation logic can produce an \emph{invalid} output for a \emph{valid} input without an explicit failure signal. Here, an input is said to be \emph{valid} if it satisfies the relevant quantum semantics and does not violate the documentation, while an output is \emph{invalid} if it violates the relevant quantum semantics or documentation. In this analysis, quantum semantics also encompass applicable software semantics. QuSema then turns each suspected violation into a testable defect hypothesis and validates it through execution. This process consists of candidate search, defect hypothesis generation, and hypothesis verification.

\textbf{Candidate Search.}
{As shown on the left of Fig.~\ref{fig:overview}(b), QuSema searches each contextual unit for semantic violations. Using its quantum knowledge, the LLM applies the source-level semantic criterion to examine implementation logic. When additional implementation context is required, the LLM follows call relations to inspect relevant definitions and code segments, and execute exploratory code to investigate suspected inconsistencies. QuSema records plausible semantic violations as defect candidates and continues the search until no further candidates are identified. Each candidate then proceeds independently through the remaining workflow.}

\textbf{Defect Hypothesis Generation.}
For each retained candidate, QuSema converts the suspected semantic violation into a testable defect hypothesis, as shown on the upper right of Fig.~\ref{fig:overview}(b). 
{Each defect hypothesis specifies the expected semantics, the implementation logic that potentially violates them, and the incorrect behavior expected under a valid triggering input.}
QuSema then constructs executable test code that directly exercises the target code segment under the identified condition. The resulting hypothesis provides a concrete prediction for subsequent verification.

\textbf{Hypothesis Verification.}
As shown on the lower right of Fig.~\ref{fig:overview}(b), QuSema executes the generated test, compares the observed behavior with the erroneous behavior predicted by the defect hypothesis, and checks whether the claimed semantic violation is justified. If the two do not match or the claimed violation is unjustified, the LLM revisits the triggering condition, test code, and predicted behavior, revises the hypothesis when needed, and repeats the verification. A candidate is retained as a semantic defect only when the predicted violation is reproduced through execution. Otherwise, it is discarded.

Stage II progresses from semantic constraints on implementation behavior to executable evidence of their violation. This process addresses \textbf{(C1)} by incorporating quantum semantic knowledge into source-level correctness reasoning.

\subsection{Library API Triggering and Bug Validation}
\label{sec:design-triggering}

The candidate defects from Stage II may either be exposed through valid use of documented library APIs or remain confined to internal implementation logic. The latter do not constitute silent bugs considered in this work. To distinguish these cases, Stage III of QuSema determines whether a verified source-level defect can be reproduced through a documented library API, thereby addressing~\textbf{(C2)}, as illustrated in Fig.~\ref{fig:overview}(c). This stage consists of library API trigger search, which identifies an API and valid input that may expose the defect, and bug qualification, which verifies whether the semantic deviation is reproduced at the API level and determines the final classification.

\textbf{Library API Trigger Search.}
As shown on the left of Fig.~\ref{fig:overview}(c), 
{QuSema searches for documented library APIs that can reach the verified defect. The LLM constructs valid inputs that may reproduce the same erroneous behavior under the triggering condition identified in Stage II, using exploratory execution when needed.} QuSema supplements this search with the reverse chain shown in the Auxiliary component of Fig.~\ref{fig:overview}(c). Starting from the smallest code segment containing the defect, the reverse chain performs a bounded search over CodeQL-derived caller relations~\cite{codeql} toward documented library APIs. For Qiskit Rust code, PyO3 bindings connect Rust implementations to the corresponding Python classes or methods. These paths identify possible API entries and guide trigger construction. An API and valid input that may reproduce the defect are retained as a trigger candidate. If no valid trigger is found, the candidate is marked \textit{not bug} and is not reported.

\textbf{Bug Qualification.}
As shown on the right of Fig.~\ref{fig:overview}(c), QuSema determines whether a trigger candidate reproduces the semantic deviation identified in Stage II. For each candidate, QuSema formulates a trigger hypothesis specifying the library API call, input, and expected erroneous behavior, then verifies it using the procedure in Sec.~\ref{sec:design-exploration}. A candidate whose predicted behavior cannot be reproduced is classified as \textit{not bug}. For a reproduced defect, QuSema checks the triggering input against the API documentation. Cases arising from inputs explicitly permitted by the documented API are classified as \textit{bug}, while cases involving valid inputs whose support is not established by the documentation are retained as \textit{potential bug} for developer assessment.

This stage determines whether an internal defect can be exposed through valid library API use. The final classification distinguishes true bugs from potential bugs according to whether the triggering input is permitted by the API documentation.

\section{Experimental Setup}
\label{sec:experimental-setup}

We evaluate the effectiveness and
utility of QuSema through the following three research questions:

\noindent\textbf{RQ1:} \emph{How does QuSema perform compared with general
coding agents and across LLM backends?}

\noindent\textbf{RQ2:} \emph{How do individual components of QuSema contribute
to its effectiveness?}

\noindent\textbf{RQ3:} \emph{Can QuSema discover previously unknown bugs in
quantum libraries?}

\textbf{Subjects and Benchmark.}
We evaluate QuSema on Qiskit and PennyLane. For RQ1 and RQ2, we construct a benchmark of 20 maintainer-confirmed historical silent bugs, with 10 from each library, and reproduce each bug on its corresponding buggy version. The benchmark covers 10 functionality types, as summarized in Table~\ref{tab:benchmark-details}. In particular, 10 of these bugs, evenly split between Qiskit and PennyLane, lack comparable implementations in the other library and therefore cannot be directly checked by differential testing. Metamorphic testing is also insufficient for these cases. For PennyLane~\#9361~\cite{pennylane9361} and Qiskit~\#12429~\cite{qiskit12429}, equivalent circuit transformations do not guarantee unchanged resource estimates. For the remaining eight bugs, equivalent circuit variants can be constructed, but their execution results do not expose the defects.
For RQ1 and RQ2, each benchmark instance centers on one historical bug. Issue reports, known triggering inputs, fixes, and regression tests are withheld from all methods. RQ1 and RQ2 evaluate defect identification and bug validation under controlled inputs, while RQ3 examines repository-scale discovery. For RQ3, we apply QuSema to the full source code of Qiskit~2.4.1~\cite{qiskitdocs} and PennyLane~0.45.0~\cite{pennylanedocs}. Table~\ref{tab:library-code-segments} reports the size, documented APIs, and extracted code segments of the two library versions.

\begin{table}[t]
\caption{Historical silent bug benchmark covering 20 maintainer-confirmed bugs in Qiskit and PennyLane. Comparable functionality indicates whether cross-library comparison is available.}
\label{tab:benchmark-details}
\centering
\footnotesize
\setlength{\tabcolsep}{1pt}
\renewcommand{\arraystretch}{0.9}
\begin{tabularx}{\linewidth}{@{}lXllll@{}}
\toprule
Category & Functionality type & Library & Issue & Version & Comparable functionality \\
\midrule
\multirow{6}{*}{\shortstack[l]{Quantum\\programming\\language}} & Semantic queries & PennyLane & \#4842~\cite{pennylane4842} & 0.33.1 & Unavailable \\
 & Semantic queries & Qiskit & \#14107~\cite{qiskit14107} & 2.2.0 & Unavailable \\
 & Construction & PennyLane & \#6359~\cite{pennylane6359} & 0.38.0 & Available \\
 & Construction & Qiskit & \#15465~\cite{qiskit15465} & 2.4.1 & Available \\
 & Specialized quantum algorithms & PennyLane & \#9521~\cite{pennylane9521} & 0.45.0 & Available \\
 & Specialized quantum algorithms & Qiskit & \#12093~\cite{qiskit12093} & 1.0.2 & Available \\
\midrule
\multirow{10}{*}{Compilation} 
 & Compilation workflow orchestration & PennyLane & \#8048~\cite{pennylane8048} & 0.42.3 & Unavailable \\
& Compilation workflow orchestration & Qiskit & \#14329~\cite{qiskit14329} & 2.1.1 & Unavailable \\
 & Structural placement decisions & PennyLane & \#9009~\cite{pennylane9009} & 0.42.3 & Unavailable \\
 & Structural placement decisions & Qiskit & \#12157~\cite{qiskit12157} & 1.3.2 & Unavailable \\
 & Resource estimation & PennyLane & \#9361~\cite{pennylane9361} & 0.45.0 & Unavailable \\
 & Resource estimation & Qiskit & \#12429~\cite{qiskit12429} & 1.0.2 & Unavailable \\
  & Representation conversion and interoperability & PennyLane & \#6978~\cite{pennylane6978} & 0.40.0 & Unavailable \\
 & Representation conversion and interoperability & Qiskit & \#10570~\cite{qiskit10570} & 1.0.2 & Unavailable \\
 & Transformation and synthesis & PennyLane & \#8746~\cite{pennylane8746} & 0.45.0 & Available \\
 & Transformation and synthesis & Qiskit & \#11974~\cite{qiskit11974} & 1.0.2 & Available \\
\midrule
\multirow{4}{*}{\shortstack[l]{Execution\\environment}} 
 & Execution and device modeling & PennyLane & \#8366~\cite{pennylane8366} & 0.42.3 & Available \\
& Execution and device modeling & Qiskit & \#13047~\cite{qiskit13047} & 1.0.2 & Available \\
 & Measurement and result processing & PennyLane & \#7344~\cite{pennylane7344} & 0.38.0 & Available \\
 & Measurement and result processing & Qiskit & \#11393~\cite{qiskit11393} & 2.4.1 & Available \\
\bottomrule
\end{tabularx}
\end{table}

\begin{table}[t]
\centering
\caption{Library statistics for RQ3. APIs count entries listed in the official documentation. Library LoC excludes comments and blank lines, while segment lengths include them and exclude attached context.}
\label{tab:library-code-segments}
\small
\setlength{\tabcolsep}{4pt}
\renewcommand{\arraystretch}{0.8}
\begin{tabular}{@{}ccccccc@{}}
\toprule
Library & Version & LoC & APIs & Code segments & Min. lines & Max. lines \\
\midrule
Qiskit & 2.4.1 & 169,750 & 899 & 8,949 & 1 & 908 \\
PennyLane & 0.45.0 & 104,172 & 1,270 & 7,786 & 2 & 446 \\
\bottomrule
\end{tabular}
\end{table}

\textbf{Configurations.}
For RQ1, we compare QuSema with Claude Code~\cite{anthropicclaudecode} using Opus~5 (High)~\cite{anthropic2026opus5} and DeepSeek Harness (DSH)~\cite{deepseek2026dsh} using DeepSeek-V4.1-Flash (High)~\cite{deepseek2026v41flash}. 
Two QuSema configurations are instantiated with Opus~5 (High) and DeepSeek-V4.1-Flash (High), respectively. Both QuSema configurations use the same workflow, prompts, tools, and execution constraints, while Claude Code and DSH retain their native workflows and tools. Both baselines are instructed to search until they identify no further silent bugs in the target source file, then report all findings in one final response. All configurations aim to identify silent bugs triggerable through valid library API use. Each QuSema model request permits up to 32,000 generated tokens, with no task-level limit on turns or cumulative token usage.
For RQ2, we use QuSema with Opus~5 (High) in two controlled comparisons. First, we evaluate candidate search under three configurations: a base QuSema configuration that analyzes complete source files without the source-level semantic oracle, a configuration with Contextual Unit Preparation, and one that additionally uses the oracle. This comparison measures their contributions to defect identification. Second, starting from the defects identified by the third configuration, we compare the subsequent workflow with and without hypothesis verification in Stages II and III. This comparison assesses the role of verification in correcting hypotheses and filtering candidates. The benchmark, environment, execution budget, and other settings remain fixed within each comparison. For RQ3, we use the complete QuSema configuration with DeepSeek-V4.1-Flash (Max).

\textbf{Procedure and Metrics.}
For RQ1, both QuSema and the baselines target the smallest source unit identified from the issue report of each bug, i.e., the complete source file containing the corresponding root cause. We compare QuSema with the baselines in terms of bug relocation count and total benchmark cost, both averaged over 3 independent runs. Concretely, a bug is \emph{relocated} when the identified defect is reproduced through valid library API use, with executable evidence and a correct explanation of the invalid output. 
QuSema partitions each source file into code segments and analyzes them independently. Therefore, for QuSema, we evaluate bug relocation on the code segments containing the root causes of the benchmark bugs, since analyzing these segments yields the same relocation count as analyzing the complete source files. We estimate the total benchmark cost for QuSema in each run by combining the measured total cost of the 20 root cause code segments with the estimated cost of all remaining segments. For each run, we independently and randomly sample 10 remaining code segments and extrapolate their mean per-segment cost to all remaining segments. For both baselines, we run each agent on the complete source file corresponding to each benchmark bug. In each run, we count the bugs that satisfy the relocation criterion and obtain the total benchmark cost by summing the measured costs across all benchmark cases. 
We also count distinct candidates across the root-cause code segments in the benchmark and record how many are verified, have library API triggers, and appear in final reports to examine the effects of hypothesis verification and API triggering in the QuSema workflow.
For RQ2, we assess component contributions using defect identification counts in the first comparison and bug relocation counts in the second. Specifically, a defect is \emph{identified} when a candidate finding matches the ground truth implementation defect and correctly explains how a valid input produces an output that violates quantum semantics or documented behavior. We apply the relocation criterion defined for RQ1. Since only the root cause code segments contribute to the counts for the target benchmark bugs, we analyze these segments for all configurations except the base configuration, which uses the complete source files containing them. For each configuration, we conduct three independent runs and report the per-run counts and the mean across runs.
For RQ3, QuSema is used to analyze code segments from entire libraries until 40 previously unknown bugs are identified. We report each unique finding to maintainers and record its status as \emph{fixed}, \emph{confirmed}, \emph{pending}, or \emph{rejected}.

\section{Evaluation Results}
\label{sec:evaluation}

This section answers the three research questions in Sec.~\ref{sec:experimental-setup}. We compare QuSema with general coding agents on the historical benchmark, examine the contributions of contextual units and the source-level semantic oracle to defect identification, and assess hypothesis verification in bug qualification. We then report previously unknown bugs found in Qiskit and PennyLane.

\subsection{RQ1: Historical Bug Relocation}

\begin{table}[t]
\caption{Bug relocation and benchmark cost on the historical benchmark. Runs 1--3 report manually reviewed counts. Mean and Mean cost are averages across three runs. Benchmark costs for both QuSema configurations are extrapolated.}
\label{tab:rq1-relocation}
\centering
\small
\setlength{\tabcolsep}{4pt}
\renewcommand{\arraystretch}{0.8}
\begin{tabular}{@{}lrrrrr@{}}
\toprule
Configuration & Run 1 & Run 2 & Run 3 & Mean & Mean cost (USD) \\
\midrule
QuSema + Opus 5 & 16 & 17 & 16 & 16.33 & 2,779 \\
QuSema + DeepSeek-V4.1-Flash & 14 & 15 & 15 & 14.67 & 99 \\
Claude Code + Opus 5 & 10 & 8 & 6 & 8.00 & 173 \\
DSH + DeepSeek-V4.1-Flash & 7 & 8 & 6 & 7.00 & 2.4 \\
\bottomrule
\end{tabular}
\end{table}

\begin{table}[t]
\caption{Candidate outcomes from QuSema's analysis of the benchmark root cause code segments.}
\label{tab:rq1-candidates}
\centering
\small
\setlength{\tabcolsep}{4pt}
\renewcommand{\arraystretch}{0.8}
\begin{tabular}{@{}llrrrr@{}}
\toprule
Configuration & Run & Candidates & Verified defects & API triggers & Final reports \\
\midrule
QuSema + Opus 5 & 1 & 42 & 42 & 42 & 42 \\
& 2 & 41 & 41 & 40 & 40 \\
& 3 & 38 & 38 & 37 & 37 \\
\midrule
QuSema + DeepSeek-V4.1-Flash & 1 & 44 & 44 & 44 & 44 \\
& 2 & 42 & 42 & 42 & 42 \\
& 3 & 47 & 46 & 46 & 46 \\
\bottomrule
\end{tabular}
\end{table}

We compare two QuSema configurations with Claude Code and DSH on the historical bug benchmark. As shown in Table~\ref{tab:rq1-relocation}, QuSema with Opus~5 achieves the highest mean bug relocation count (16.33), followed by QuSema with DeepSeek (14.67), Claude Code (8.00), and DSH (7.00). DeepSeek-based QuSema achieves a substantially higher mean bug relocation count than Claude Code at a lower estimated mean benchmark cost (\$99 versus \$173). Using a less expensive LLM backend, QuSema can balance effectiveness and cost better than Claude Code.

From the root cause code segments, QuSema produces 38--47 candidates and 37--46 final reports per run (Table~\ref{tab:rq1-candidates}). Defect hypothesis verification and library API trigger search exclude 0--1 candidates per run because verification does not support the claimed bug or no reachable library API trigger is found. Among the reports beyond the benchmark targets, four correspond to additional historical bugs documented in existing issue reports. Manual review finds the remaining reports consistent with our bug criteria, although they have not yet been submitted for maintainer validation. These results show that QuSema can identify findings beyond the benchmark targets while filtering unsupported or non-triggerable candidates through Stages II and III.

\begin{rqanswerbox}
\textbf{Answer to RQ1.}
QuSema achieves mean bug relocation counts of 16.33 with Opus and 14.67 with DeepSeek, compared with 8.00 for Claude Code and 7.00 for DSH. DeepSeek-based QuSema achieves a substantially higher mean relocation count than Claude Code at a lower estimated mean benchmark cost.
\end{rqanswerbox}

\subsection{RQ2: Component Contributions}

We conduct two controlled comparisons of QuSema components, as shown in Fig.~\ref{fig:rq2-ablation}. The first evaluates Contextual Unit Preparation and the Source-level Semantic Oracle for defect identification, while the second evaluates hypothesis verification for bug relocation.

In the first comparison, the base configuration analyzing complete source files identifies 5--6 defects across three runs, averaging 5.67. Contextual Unit Preparation raises the mean to 14.00, with 14 defects identified in each run, and adding the Source-level Semantic Oracle raises it to 16.33. The larger gain from Contextual Unit Preparation suggests that reducing unrelated context helps focus defect analysis. The oracle's additional gain indicates that semantic evidence helps determine correctness when observed behavior alone is insufficient. The two components address complementary sources of failure: Contextual Unit Preparation focuses the analysis on relevant implementation context, while the Source-level Semantic Oracle provides evidence for distinguishing semantic defects from permissible implementation choices.

Hypothesis verification improves the correctness of final reports. It corrects matrix traces and an overly narrow description of triggering conditions for PennyLane~\#8746~\cite{pennylane8746}, and the expected resource count for \#9361~\cite{pennylane9361}. Without verification, these errors persist until API trigger construction corrects them. However, trigger construction introduces three errors across two Qiskit~\#15465~\cite{qiskit15465} reports and one PennyLane~\#7344~\cite{pennylane7344} report. Their descriptions of the erroneous behavior conflict with the execution results. Without verification, these inconsistencies persist in the final reports, preventing the reported evidence from establishing the claimed bugs. Full QuSema corrects these reports, raising mean relocation from 15.33 to 16.33.

\begin{rqanswerbox}
\textbf{Answer to RQ2.}
Contextual Unit Preparation and the Source-level Semantic Oracle increase mean defect identification from 5.67 to 14.00 and 16.33, respectively. Hypothesis verification corrects errors that persist through subsequent analysis, increasing mean bug relocation from 15.33 to 16.33.
\end{rqanswerbox}

\begin{figure}[t]
\centering
\includegraphics[width=\linewidth]{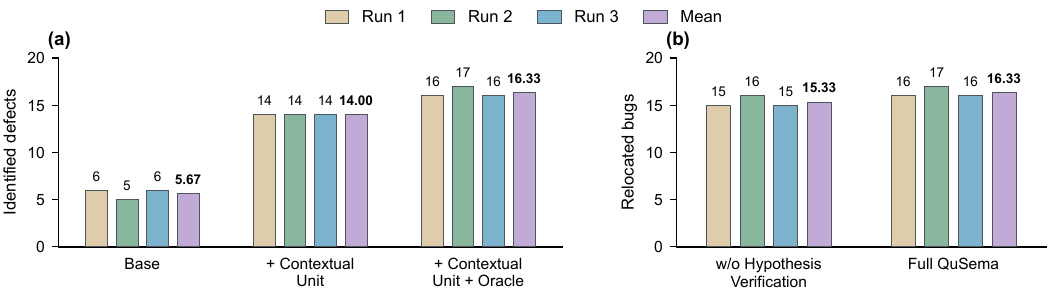}
\caption{Component contributions with Opus~5 (High). (a) Candidates matching historical defects as Contextual Unit Preparation and the Source-level Semantic Oracle are added. (b) Bugs satisfying the relocation criterion with and without hypothesis verification in Stages II and III. Each group reports the counts from Runs 1--3 and their mean across the three runs.}
\Description{Two bar charts on the same row. Each group contains four bars for Runs 1--3 and their mean. The left chart shows identified historical defects. Base identifies 6, 5, and 6, with mean 5.67. Adding Contextual Unit Preparation yields 14, 14, and 14, with mean 14.00. Further adding the Source-level Semantic Oracle yields 16, 17, and 16, with mean 16.33. The right chart shows relocated historical bugs. Without hypothesis verification, the counts are 15, 16, and 15, with mean 15.33. Full QuSema achieves 16, 17, and 16, with mean 16.33.}
\label{fig:rq2-ablation}
\end{figure}

\subsection{RQ3: Previously Unknown Bugs}

QuSema with DeepSeek-V4.1-Flash (Max) is used to analyze the source code of Qiskit~2.4.1 and PennyLane~0.45.0 for previously unknown bugs. It analyzes 68,687 lines and 2,753 code segments in Qiskit, and 72,262 lines and 2,269 code segments in PennyLane. QuSema reports 40 previously unknown bugs, all confirmed by the maintainers. Among them, 30 are silent bugs that return invalid outputs without explicit failure signals, while the other 10 cause crashes or exceptions on valid inputs. Tables~\ref{tab:bugs-qiskit} and~\ref{tab:bugs-pennylane} summarize these findings by API location and bug type, where \sem{} denotes a silent bug and \crash{} a crash or exception.

The confirmed bugs span circuit transformation, symbolic computation, commutativity, measurement, resource estimation, serialization, and execution. Among the 30 silent bugs, 5 affect PennyLane functionalities without directly comparable APIs in the evaluated Qiskit SDK, preventing direct cross-library comparison. These cases also lack applicable circuit-based metamorphic relations. Specifically, Bosonic operator multiplication, measurement condition comparison, and download concurrency are not exercised by equivalent-circuit variants, while lost wire count information and parameter trainability can remain incorrect even when circuit execution is unchanged.

\begin{figure}
\centering
\includegraphics[width=\linewidth]{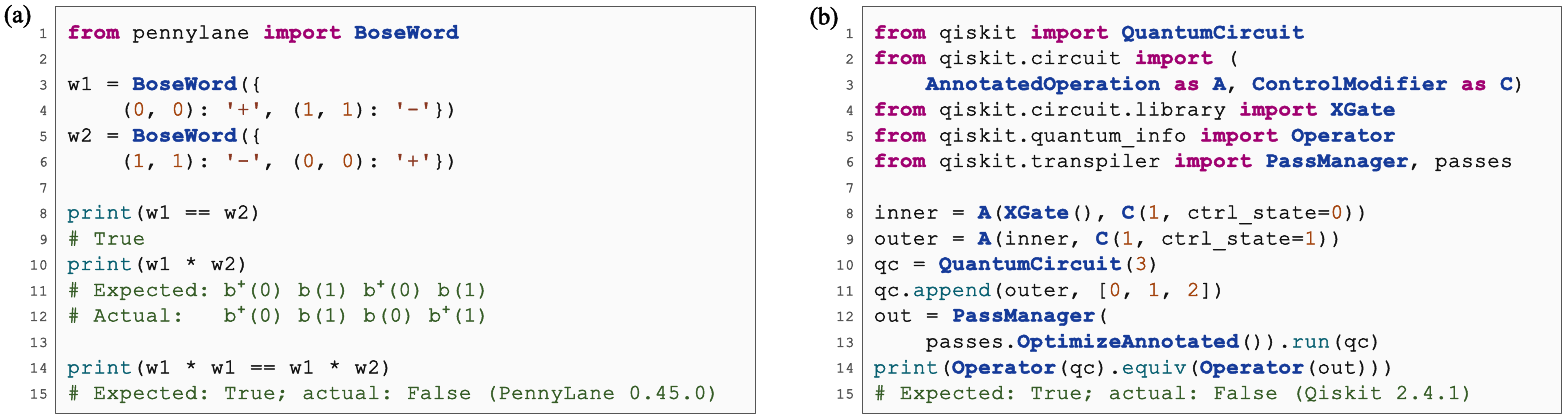}
\caption{Silent bugs discovered by QuSema: (a) incorrect bosonic operator multiplication in PennyLane \#9650~\cite{pennylane9650} and (b) incorrect control state composition in Qiskit \#16410~\cite{qiskit16410}.}
\label{fig:rq3-nested-control}
\Description{Two Python reproducers. In panel (a), BoseWord objects constructed from identical dictionary entries in different insertion orders compare equal, but multiplication returns different products. In panel (b), OptimizeAnnotated combines nested controls incorrectly and changes the circuit unitary.}
\end{figure}

Figure~\ref{fig:rq3-nested-control} shows two representative silent bugs. In Fig.~\ref{fig:rq3-nested-control}(a), PennyLane's \texttt{BoseWord} multiplication (\#9650~\cite{pennylane9650}) produces incorrect results for equivalent operands constructed from identical dictionary entries in different insertion orders. The implementation pairs sorted dictionary keys with values retained in insertion order, which can exchange creation and annihilation operators during multiplication. This symbolic operator functionality has no direct counterpart in the evaluated Qiskit SDK and no applicable circuit-based metamorphic relation. Figure~\ref{fig:rq3-nested-control}(b) shows Qiskit~\#16410~\cite{qiskit16410} in the \texttt{OptimizeAnnotated} pass. The implementation combines nested control modifiers in the wrong order. For an inner open control and an outer closed control, it produces control state~2 instead of~1, changing the control condition and thereby violating circuit semantics without raising an error. The maintainer response~\cite{qiskit16410} highlighted that QuSema pinpointed the root cause:

\begin{center}
\setlength{\fboxsep}{4pt}
\fbox{%
\begin{minipage}{0.98\linewidth}
\emph{``I have pushed a fix in \#16428~\cite{qiskit16428}, but here is a general question for you: would you be interested in submitting PRs for some of the issues that you are discovering (especially that you already do the work of pinpointing the exact root cause for each problem)?''}
\end{minipage}%
}
\end{center}

The response indicates that the root cause information reported by QuSema can provide developers with actionable guidance for subsequent bug diagnosis and repair.

\begin{rqanswerbox}
\textbf{Answer to RQ3.} QuSema finds 40 previously unknown bugs confirmed by maintainers, including 5 silent bugs that existing differential and metamorphic testing cannot detect.
\end{rqanswerbox}

\begin{table}[tbp]
\centering
\begin{minipage}[t]{0.485\linewidth}
\centering
\caption{Previously unknown bugs in Qiskit.}
\label{tab:bugs-qiskit}
\footnotesize
\setlength{\tabcolsep}{1pt}
\renewcommand{\arraystretch}{0.9}
\begin{tabularx}{\linewidth}{@{}rl>{\raggedright\arraybackslash}Xl@{\hspace{8pt}}l@{}}
\toprule
\# & Issue & API Location & Type & Status \\
\midrule
1 & \#16271~\cite{qiskit16271} & \texttt{transpiler.\allowbreak{}passes} & \sem & Confirmed\\
2 & \#16205~\cite{qiskit16205} & \texttt{transpiler.\allowbreak{}passes} & \sem & Confirmed\\
3 & \#16185~\cite{qiskit16185} & \texttt{transpiler.\allowbreak{}passes} & \sem & Confirmed\\
4 & \#16166~\cite{qiskit16166} & \texttt{converters} & \sem & Confirmed\\
5 & \#16377~\cite{qiskit16377} & \texttt{transpiler.\allowbreak{}passes} & \sem & Confirmed\\
6 & \#16259~\cite{qiskit16259} & \texttt{circuit} & \sem & Confirmed\\
7 & \#16260~\cite{qiskit16260} & \texttt{circuit} & \sem & Confirmed\\
8 & \#16262~\cite{qiskit16262} & \texttt{circuit} & \sem & Confirmed\\
9 & \#16263~\cite{qiskit16263} & \texttt{circuit} & \sem & Confirmed\\
10 & \#16161~\cite{qiskit16161} & \texttt{transpiler.\allowbreak{}passes} & \sem & Confirmed\\
11 & \#16164~\cite{qiskit16164} & \texttt{circuit} & \sem & Confirmed\\
12 & \#16410~\cite{qiskit16410} & \texttt{transpiler.\allowbreak{}passes} & \sem & Confirmed\\
13 & \#16411~\cite{qiskit16411} & \texttt{transpiler.\allowbreak{}passes} & \sem & Confirmed\\
14 & \#16184~\cite{qiskit16184} & \texttt{circuit.\allowbreak{}library} & \crash & Confirmed\\
15 & \#16173~\cite{qiskit16173} & \texttt{primitives.\allowbreak{}containers} & \sem & Confirmed\\
16 & \#16168~\cite{qiskit16168} & \texttt{circuit.\allowbreak{}library} & \sem & Confirmed\\
17 & \#16190~\cite{qiskit16190} & \texttt{result} & \sem & Confirmed\\
18 & \#16412~\cite{qiskit16412} & \texttt{circuit} & \crash & Confirmed\\
19 & \#16413~\cite{qiskit16413} & \texttt{circuit.\allowbreak{}library} & \sem & Confirmed\\
20 & \#15841~\cite{qiskit15841} & \texttt{transpiler.\allowbreak{}passes} & \crash & Confirmed\\
\bottomrule
\end{tabularx}
\end{minipage}
\hfill
\begin{minipage}[t]{0.485\linewidth}
\centering
\caption{Previously unknown bugs in PennyLane.}
\label{tab:bugs-pennylane}
\footnotesize
\setlength{\tabcolsep}{1pt}
\renewcommand{\arraystretch}{0.9}
\begin{tabularx}{\linewidth}{@{}rl>{\raggedright\arraybackslash}Xl@{\hspace{8pt}}l@{}}
\toprule
\# & Issue & API Location & Type & Status \\
\midrule
1 & \#9612~\cite{pennylane9612} & \texttt{devices} & \crash & Confirmed\\
2 & \#9613~\cite{pennylane9613} & \texttt{devices.\allowbreak{}preprocess} & \crash & Confirmed\\
3 & \#9617~\cite{pennylane9617} & \texttt{estimator} & \crash & Confirmed\\
4 & \#9633~\cite{pennylane9633} & \texttt{ops.\allowbreak{}op\_math} & \crash & Confirmed\\
5 & \#9652~\cite{pennylane9652} & \texttt{debugging} & \crash & Confirmed\\
6 & \#9610~\cite{pennylane9610} & \texttt{devices} & \sem & Confirmed\\
7 & \#9631~\cite{pennylane9631} & \texttt{ops.\allowbreak{}op\_math} & \sem & Confirmed\\
8 & \#9632~\cite{pennylane9632} & \texttt{ops.\allowbreak{}functions} & \sem & Confirmed\\
9 & \#9650~\cite{pennylane9650} & \texttt{bose} & \sem & Confirmed\\
10 & \#9669~\cite{pennylane9669} & \texttt{devices} & \sem & Confirmed\\
11 & \#9622~\cite{pennylane9622} & \texttt{ops.\allowbreak{}functions} & \sem & Confirmed\\
12 & \#9623~\cite{pennylane9623} & \texttt{ops.\allowbreak{}functions} & \sem & Confirmed\\
13 & \#9653~\cite{pennylane9653} & \texttt{devices} & \sem & Confirmed\\
14 & \#9670~\cite{pennylane9670} & \texttt{measurements} & \sem & Confirmed\\
15 & \#9616~\cite{pennylane9616} & \texttt{estimator} & \sem & Confirmed\\
16 & \#9618~\cite{pennylane9618} & \texttt{estimator} & \crash & Confirmed\\
17 & \#9609~\cite{pennylane9609} & \texttt{data} & \sem & Confirmed\\
18 & \#9614~\cite{pennylane9614} & \texttt{debugging} & \sem & Confirmed\\
19 & \#9601~\cite{pennylane9601} & \texttt{concurrency.\allowbreak{}executors} & \crash & Confirmed\\
20 & \#9606~\cite{pennylane9606} & \texttt{data} & \sem & Confirmed\\
\bottomrule
\end{tabularx}
\end{minipage}
\end{table}

\section{Discussion and Threats to Validity}
\label{sec:threats-to-validity}

\textbf{Discussion.}
QuSema localizes semantic deviations to responsible code segments and reproduces them through documented library APIs. The resulting reports provide root cause information and user-level triggers that support diagnosis and repair, as also reflected in the maintainer feedback in RQ3. Differential and metamorphic testing rely on comparable implementations or predictable execution relations, whereas QuSema examines implementation logic against quantum semantics and documentation. This allows it to detect confirmed silent bugs in RQ3 for which such relations are unavailable or uninformative.

\textbf{Construct Validity.}
QuSema relies on LLM knowledge of quantum computing and software to identify defects. This knowledge may be incomplete, and hallucinations~\cite{kalai2026hallucinations} may cause incorrect semantic judgments. Hypothesis verification tests these judgments against execution evidence but cannot guarantee reliable detection. QuSema also uses documentation as semantic evidence of intended behavior without independently establishing whether the documentation itself is correct. Documentation errors~\cite{xu2025documentation} may therefore affect bug assessment and require maintainer review.

\textbf{Internal Validity.}
LLM outputs vary across runs~\cite{ouyang2025nondeterminism} and model versions. We repeat each configuration three times under the same experimental settings. Candidate reports are reviewed manually, so assessment may involve subjective judgment. The checks described in Sec.~5 help limit this subjectivity. RQ1 compares QuSema with Claude Code and DSH as complete configurations, so performance differences cannot be attributed to agent design alone. The 10 code segments independently and randomly sampled in each run may not capture cost variation across the remaining 588 segments. Public benchmark issues and fixes may have appeared in LLM training data despite being withheld from the evaluated methods. RQ3 reduces this threat by evaluating previously unknown bugs, although prior exposure to the underlying library source code cannot be excluded.

\textbf{External Validity.}
The historical benchmark contains 20 bugs from Qiskit and PennyLane, and practical discovery uses Qiskit~2.4.1 and PennyLane~0.45.0. The results may not generalize to closed-source or poorly documented libraries, or to different implementation languages and binding systems. Applying QuSema may require changes to code segmentation, call relation analysis, and documentation association. QuSema targets API-local silent bugs and does not directly address bugs arising solely from interactions among individually correct APIs.

\section{Related Work}
\label{sec:related-work}

\textbf{Quantum software correctness and testing.}
Research on quantum software correctness spans language design, formal verification, static analysis, and testing. At the program level, Silq~\cite{bichsel2020silq} supports safe automatic uncomputation through its language and type system, while VOQC~\cite{hietala2021voqc} verifies that circuit optimizations preserve quantum semantics. LintQ~\cite{paltenghi2024lintq} uses quantum-specific abstractions to statically detect problems in Qiskit programs. Beyond program-level correctness, quantum software testing addresses library reliability. Differential and metamorphic testing detect bugs through cross-library comparisons or semantics-preserving transformations, while KQFuzz~\cite{kqfuzz} uses codebase knowledge to guide LLM-based test generation. QuSema instead uses quantum semantics and documentation as source-level correctness evidence for auditing library logic, without requiring predefined relations between executions.

\textbf{LLM-based software analysis and semantic auditing.}
LLMs support repository-level issue resolution, test and oracle generation, and source auditing. For general software engineering tasks, SWE-agent~\cite{yang2024sweagent} and RepairAgent~\cite{bouzenia2025repairagent} use repository tools to resolve software issues. CoverUp~\cite{coverup} uses coverage and execution feedback to generate tests, while Doc2OracLL~\cite{hossain2025doc2oracll} uses documentation for test oracle generation. For source-level auditing, RepoAudit~\cite{guo2025repoaudit} validates data flow and path conditions during repository exploration, whereas RFCAudit~\cite{zheng2025rfcaudit} checks implementations against RFC specifications and retrieves related source code as needed. QuSema combines documentation and quantum semantics to audit quantum library source code and validate whether identified defects can be exposed through documented library APIs.

\section{Conclusion}
\label{sec:conclusion}

This paper presents QuSema, an agentic approach for detecting silent bugs in quantum libraries using source-level semantic evidence. QuSema analyzes implementation logic against quantum semantics and documentation to identify candidate defects. Library API triggering establishes whether these defects produce invalid outputs under valid library use. This approach does not depend on cross-library counterparts or predefined metamorphic relations. Applied to Qiskit and PennyLane, QuSema discovers 40 previously unknown bugs.

\bibliographystyle{ACM-Reference-Format}
\bibliography{references}

\end{document}